\documentclass[12pt]{article}
\usepackage{epsfig}

\newcommand{\bea}{\begin{eqnarray}}
\newcommand{\eea}{\end{eqnarray}}
\newcommand{\mathbold}[1]{\mbox{\boldmath $\bf#1$}}

\begin{document}
\setlength{\baselineskip}{0.7cm}

\begin{titlepage}
\null
\begin{flushright}
KEK Preprint 2004-1\\
KEK-TH-947
\end{flushright}
\vskip 1cm
\begin{center}
{\LARGE\bf 
Higgs pair production at a linear $\protect \mathbold{e^+e^-}$  collider \\
\vspace{3mm}
in models with large extra dimensions 
} 

\lineskip .75em
\vskip 1.5cm

\normalsize

{\large\bf 
Nicolas Delerue
\footnote{E-Mail: nicolas@post.kek.jp}, 
Keisuke Fujii
\footnote{E-Mail: keisuke.fujii@kek.jp}
and Nobuchika Okada
\footnote{E-Mail: nobuchika.okada@kek.jp}
}

\vspace{10mm}

{\it 
High Energy Accelerator Research Organization (KEK) \\ 
1-1 Oho, Tsukuba, Ibaraki 305-0801, JAPAN} \\

\vspace{9mm}

{\it To be submitted to Physical Review Journal D, rapid communications}

\vspace{9mm}

{\bf Abstract}\\[5mm]
{\parbox{13cm}{\hspace{5mm}
%
%%%%%%%%%%%%%%%%%%%%%%%%%%%%%%%%%%%%%%%%%%%%%%%%%%%%%%%%%%%%%%%%%%
In this paper, we will derive the cross section formula for the Higgs pair 
production at a linear $e^+e^-$ collider in models with large extra dimensions
and study the feasibility of its measurement through realistic Monte Carlo
simulations.
Since the process has essentially no Standard Model background,
once produced, it will provide us with a very clean signature of
physics beyond the Standard Model.
Moreover, since the final state particles are spinless, the spin 2 of 
the intermediate virtual KK gravitons has to be conserved by the
orbital angular momentum of the Higgs pair.
This results in a very characteristic angular distribution of the final states.
Taking into account finite detector acceptance and resolutions 
as well as initial state radiation and beamstrahlung,
we demonstrate in this paper that, given a sufficiently high center 
of mass energy, the angular distribution of the Higgs pair is 
indeed measurable at the linear collider 
and will allow us to prove the spin 2 nature of the KK gravitons 
exchanged in the $s$-channel.

%%%%%%%%%%%%%%%%%%%%%%%%%%%%%%%%%%%%%%%%%%%%%%%%%%%%%%%%%%%%%%%%%%%
}}
\end{center}
\end{titlepage}

\section{Introduction}
%%%%%%%%%%%%%%%%%%%%%%%%%%%
% Main body 
%%%%%%%%%%%%%%%%%%%%%%%%%%%
Large extra dimension scenario \cite{ADD} 
provides an alternative solution to the gauge hierarchy problem, 
the huge hierarchy between the electroweak scale and 
the Planck scale, without supersymmetry. 
%In this scenario, while the original gravity scale 
%can be of order 1 TeV, the four dimensional Planck scale 
%is obtained by the arrangement of the large extra dimensional volume. 
In this scenario, while keeping the four dimensional Planck scale as it is,
we can bring the fundamental gravity scale down as low as to ${\cal O}$(1TeV),
by the arrangement of the extra dimensional volume.
For example, in the case of models with extra $\delta$ dimensions 
being compactified on a torus $T^\delta$ with a common radius $R$, 
the four dimensional Planck scale ($M_4$) is obtained 
through the Gauss law, $M_4^2 \sim M_{4+\delta}^{2+\delta} R^\delta$, 
where $M_{4+\delta}$ is the  $4+\delta$ dimensional Planck scale. 
Since all the standard model fields are assumed to be confined 
on the brane, the existence of the extra dimensions 
can be felt only through the very weak gravity interactions, 
and the models are found to be consistent with the current experiments 
even if $M_{4+\delta} \sim {\cal O}(1~\mbox{TeV})$ 
for $\delta \geq 2$ \cite{Hoyle-etal}. 
Of great interests is that this scenario is testable at colliders, 
and its collider signatures 
have been intensively investigated \cite{review}. 

In the four dimensional description, 
there is an infinite tower of Kaluza-Klein (KK) modes of graviton, 
whose masses are characterized by the compactification radius 
and integers corresponding to quantization of 
momentum in the extra dimensional directions. 
The existence of the extra dimensions can be revealed 
through processes involving the KK mode gravitons at the colliders. 
There are two important types of processes. 
One involves direct emission of the KK gravitons. 
Since KK gravitons interact with the standard model fields 
very weakly, they are regarded as  invisible stable particles. 
Such a process will thus appear as 
missing energy events at the colliders. 
The other comprises the processes mediated by virtual KK gravitons. 
In this case, new effective contact interactions will be induced 
in addition to the standard model interactions.

In this paper, we study the Higgs pair production process 
at a future linear $e^+ e^-$ collider\footnote{As related works, see \cite{Rizzo:1999qv}, \cite{KLS}  and \cite{He}, 
where the Higgs pair production at $e^+e^-$ colliders, LHC and  photon colliders, 
respectively, have been discussed.}.
This process has two interesting points to be stressed. 
Firstly, the Higgs pair production cross section at the linear collider
is highly suppressed in the standard model. 
If the cross section mediated by the virtual KK gravitons 
is large enough, it will hence be a very clean signal
of physics beyond the Standard Model. 
Secondly, the clean and well defined environment of the linear
collider allows the measurement of  the angular distribution of 
the final states, which carries information 
on the spins of the intermediate states. 
Being scalar particles, the produced pair of Higgs bosons will show 
a very characteristic angular distribution, 
since the spin 2 of the intermediate KK gravitons has to be
conserved by the orbital angular momentum of the final states. 

In the next section we will derive the cross section formula for this
process and show this explicitly.
Based on the derived formula, we will then carry out Monte Carlo simulations
in Section \ref{Sec:MC} and demonstrate that we can indeed
measure the angular distribution. 
The final section summarizes our results and concludes this paper.

\section{Theoretical Framework}

In this section, we will derive the cross section formula
for the Higgs pair production.
For simplicity, we assume the extra $\delta$ dimensions are 
compactified on the torus $T^\delta$ with a common radius $R$. 
In the four dimensional description, 
the interaction Lagrangian between the standard model fields 
and the  KK gravitons ($ G^{(\vec{n})}_{\mu \nu} $) 
or KK gravi-scalars ($H^{(\vec{n})}$) is then given by 
\cite{GRW,HLZ} 
\bea 
 {\cal L}_{int} = - \frac{1}{{\bar M}_p} 
\sum_{\vec{n}} \left( 
G^{(\vec{n})}_{\mu \nu} T^{\mu \nu} + 
H^{(\vec{n})} T^{\mu}_{\mu}   \right) ,  
 \label{interaction} 
\eea  
where $\vec{n}=(n_1,n_2,..,n_\delta)$ with $n_i$'s being integers, 
$ T_{\mu \nu}$ is the energy-momentum tensor 
of the standard model fields, and ${\bar M}_p=M_4/\sqrt{8 \pi}$ 
is the reduced four dimensional Planck scale. 
The $n$-th KK mode mass squared is characterized by 
$m_{(\vec{n})}^2= |\vec{n}|^2/R^2$. 
Since the trace of the energy-momentum tensor is proportional to 
the mass of the fields under the field equations of motion, 
we neglect processes mediated by the KK gravi-scalars 
in our study for the linear $e^+ e^-$ collider, 
and consider only the processes mediated by the KK gravitons. 

In this setup, the scattering amplitude via the $s$-channel KK
graviton exchange for a general 2-to-2 process, 
$e^{-}(p_1)+e^{+}(p_2) \rightarrow X(p_3)+Y(p_4)$, is found to be 
\bea
{\cal M} =  \sum_{\vec{n}} 
\left(-\frac{1}{\bar{M}_p^2} \right) 
 \frac{1}{s^2- m_{(\vec{n})}^2} 
T^{\mu \nu} (p_1 , p_2) T_{\mu \nu} (p_3 , p_4) 
 \label{amplitude}
\eea
in momentum space. 
Note that there is a difficulty in this process (for $\delta \geq 2$), 
namely, the ultraviolet divergence 
according to the summation over the infinite tower of the KK modes. 
Unfortunately no one knows how to regularize 
the divergence because of the lack of knowledge 
about gravity theories at energies much higher 
than the ($4+\delta$ dimensional) Planck scale. 
In the following analysis, we naively introduce an ultraviolet cutoff 
for the highest KK modes, and replace the summation by \cite{GRW}  
\bea 
\frac{4 \pi}{M_S^4} = - \frac{1}{\bar{M}_p^2} 
\sum_{\vec{n}} \frac{1}{s-m_{(\vec{n})}^2} , 
\eea
where $M_S$ is the cutoff scale naturally being 
of the order of $M_{4+\delta}$.
\footnote{
When a finite brane tension or a finite brane width is introduced, 
the ultraviolet divergence is automatically regularized, 
and the finite result is obtained \cite{BKNY,HO}. 
In this case, $M_S$ is related to the brane tension or 
the brane width.  We can thus interpret our results below 
as from models with a finite brane tension or 
finite brane width. 
} 
Our amplitude will then become 
\bea 
{\cal M} =
 \frac{4 \pi}{M_S^4}  
 T^{\mu \nu} (p_1 , p_2) T_{\mu \nu} (p_3 , p_4) . 
\eea 
We use this formula for 
the collider's center of mass energies below $M_S$.

Let us apply the framework given above to the Higgs pair production process, 
$e^+ e^- \rightarrow HH$, mediated by the virtual KK gravitons.  
Only the free parts in the (physical) Higgs Lagrangian, 
the kinetic and the mass terms, contribute to this process. 
In this process, the energy momentum tensors in the amplitude 
are explicitly given by 
\bea 
T^{\mu \nu} (p_1,p_2) 
&=&  \frac{1}{4} \bar{v}(p_2) 
 \left[  (p_1-p_2)^\mu \gamma^\nu +(p_1-p_2)^\nu \gamma^\mu  \right] 
 u(p_1) ,   \nonumber \\ 
T^{\mu \nu} (p_3,p_4) &=& 
  -p_3^\mu p_4^\nu -p_3^\nu p_4^\mu  
  +\eta^{\mu \nu} \left(  (p_3 \cdot p_4) +m_H^2 \right) , 
\eea
respectively. 
After straightforward calculations, 
the squared amplitude is found to be  
\bea 
 \sum_{spin} |{\cal M}|^2= \frac{1}{2} 
\left( \frac{4 \pi}{M_S^4} \right)^2  (t-u)^2 (t u -m_H^4)  ,
\eea 
or, in terms of the production angle $\theta$,
\bea
 \sum_{spin} |{\cal M}|^2= \frac{1}{8} 
\left( \frac{4 \pi}{M_S^4} \right)^2 
s^4 \beta^4 \sin^2\theta \cos^2\theta  ,
\eea
where $\beta = \sqrt{1 - 4 m_H^2/s}$.
The factor of $\beta^4$ indicates the D-wave ($L=2$) contribution.
The amplitude leads us to the following formula 
for the differential cross section:
\bea
\label{eqn:dsigma}
\frac{d \sigma}{d \cos\theta} (e^{+}e^{-} \rightarrow HH) 
& = & \frac{\pi }{128 M_S^8} s^3 \beta^5 \sin^2\theta \cos^2\theta . \label{eq:diffcrosssection}
\eea
The total cross section is then readily obtained as 
\bea 
\sigma(e^{+}e^{-} \rightarrow HH) 
%& = & \frac{\pi }{480 M_S^8}
% \sqrt{1-4 \frac{m_H^2}{s}} 
% \left( s^3 - 8 m_H^2 s^2 + 16 m_H^4 s  \right)  \cr
 & = & \frac{\pi }{480 M_S^8} s^3 \beta^5. \label{eq:totcrosssection}
\eea
For $s \gg m_H^2$, 
the total cross section is proportional to $s^3$ and $1/M_S^8$. 
At a linear collider with $\sqrt{s}=1~\mbox{TeV}$, 
we obtain $\sigma(e^{+}e^{-} \rightarrow HH) \sim 8.6~\mbox{fb}$ 
for $M_S=2~\mbox{TeV}$ and the higgs mass $m_H=120~\mbox{GeV}$. 
Interestingly this cross section is of the same order of 
magnitude as that of the associated higgs production cross section, 
$e^+ e^- \rightarrow Z H$ in the standard model. 

\begin{figure}
\begin{center}
\epsfig{file=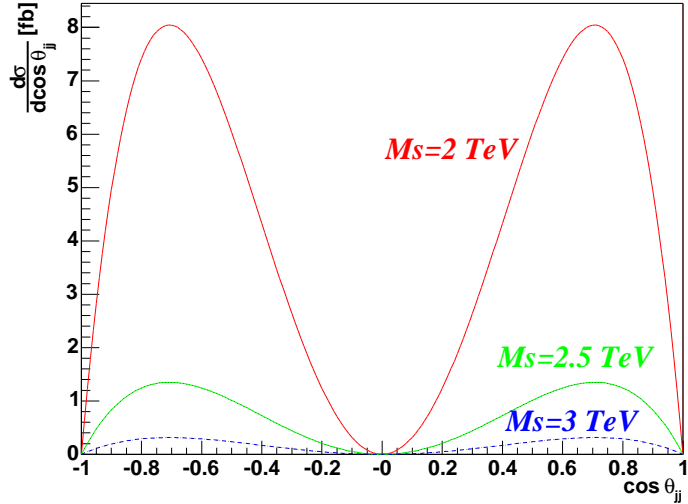, width=10cm}
\caption{
Angular dependence of the cross section 
$\frac{d\sigma(e^{+}e^{-} \rightarrow HH)}{d \cos \theta}$ [fb]  
(Eq.~\ref{eqn:dsigma}) as a function of the scattering angle ($\cos \theta$) 
for $M_S=2$, $2.5$, and 3~TeV, respectively, from top to bottom. 
}
\label{Fig:dsdcosth}
\end{center}
\end{figure}

We show a characteristic behavior of 
the angular dependence of the cross section in Fig.~\ref{Fig:dsdcosth}. 
%%%%%%%%%%%%%%%%%%%%%%%%%%%
%  Fig1
%%%%%%%%%%%%%%%%%%%%%%%%%%
\newpage
%%%%%%%%%%%%%%%%%%%%%%%%%%%
%%%%%%%%%%%%%%%%%%%%%%%%%

There are peaks in forward and backward regions.
Notice also that the differential cross section vanishes at
three points: $\cos\theta = 0$ and $\pm 1$.
This can be qualitatively understood in the following manner.
Since the initial state consists of two spin-1/2 particles,
the spin sum can not exceed 1.
There must hence be some finite orbital angular momentum
in the initial $e^+e^-$ system to make up the spin 2 of the
intermediate KK gravitons.
The orbital angular momentum is, however, perpendicular to the
momentum, which means that the intermediate KK gravitons
must have a spin component perpendicular to the beam axis.
This component, however, has to be averaged to zero because
of the rotational symmetry about the beam axis.
On the other hand, since the structure of the energy momentum tensor
of the initial state is chirality conserving,
the initial-state spin sum and hence the spin of the KK gravitons
must have a component of $\pm 1$ along the beam axis.
Since the final-state Higgs bosons are spinless,
this angular momentum of the KK gravitons has to be
conserved by their orbital angular momentum.
This implies that the final-state Higgs pair is in
the $Y_l^m = Y_2^{\pm 1}$ state with the beam axis
taken as its quantization axis.
Eq. (7) indicates that this is indeed the case.
Since $Y_2^{\pm 1}$ is odd in terms of $\cos\theta$,
the cross section has to vanish at $\cos\theta=0$.
Similarly the vanishment of the differential cross section at
$\cos\theta = \pm 1$ can be interpreted as a consequence
of the finite spin component in the beam direction:
at $\cos\theta=\pm 1$ this component must vanish
since the final-state orbital angular momentum is perpendicular
to the beam axis.
%This means that spin 0 combination of the initial state is forbidden.
%This is the result of the structure of the energy momentum tensor
%of the initial state that is chirality conserving.

\section{Monte Carlo Simulation}
\label{Sec:MC}

\subsection{Signal and Background Samples}

A Monte Carlo event generator  has been written for the Higgs pair production,
using the above cross section formula.
The generator generates 4-momenta of the final-state Higgs bosons,
taking into account the initial state radiation and beamstrahlung
and pass them to a hadronizer module 
of JLC Study Framework (JSF)\cite{Ref:JSF, Ref:JLC}.
This hadronizer is based on Pythia 6\cite{Ref:Pythia}, decays the final-state Higgs bosons
into lighter partons, and makes them parton-shower and fragment,
if necessary.
The final-state jets of particles are then passed to the JSF's quick detector simulator
module to emulate  the JLC detector response as described in \cite{Ref:JLC}. 
The charged particle tracks and calorimeter clusters are then combined to
measure energy flows from the individual jets and then used to reconstruct
the final-state Higgs bosons as described below.

The signature of the signal process is the production of 
two pairs of $b$-jets whose mass is close to the mass of the Higgs. 
Other Standard Model processes producing two pairs of jets that mimic the signal 
will be our main backgrounds.
The potential background processes thus include
$e^+e^- \to W^+W^-$, $ZZ$, $ZH$ and $b\bar{b}b\bar{b}$.
We used the physsim package \cite{Ref:JLC, Ref:physsim} and GRC4F\cite{Fujimoto:1996wj} to generate
these background events.
The package works under JSF and uses full helicity amplitudes so that the
angular correlations of the final state partons can be properly
taken into account.

In what follows, we assume a linear collider of $\sqrt{s} = 1$ TeV and
study feasibility of measuring the angular distribution for the Higgs
pair production process for
$M_S=2~\mbox{TeV}$ and $m_H=120~\mbox{GeV}$.

\subsection{Event Selection}

\begin{figure}
\begin{center}
\epsfig{file=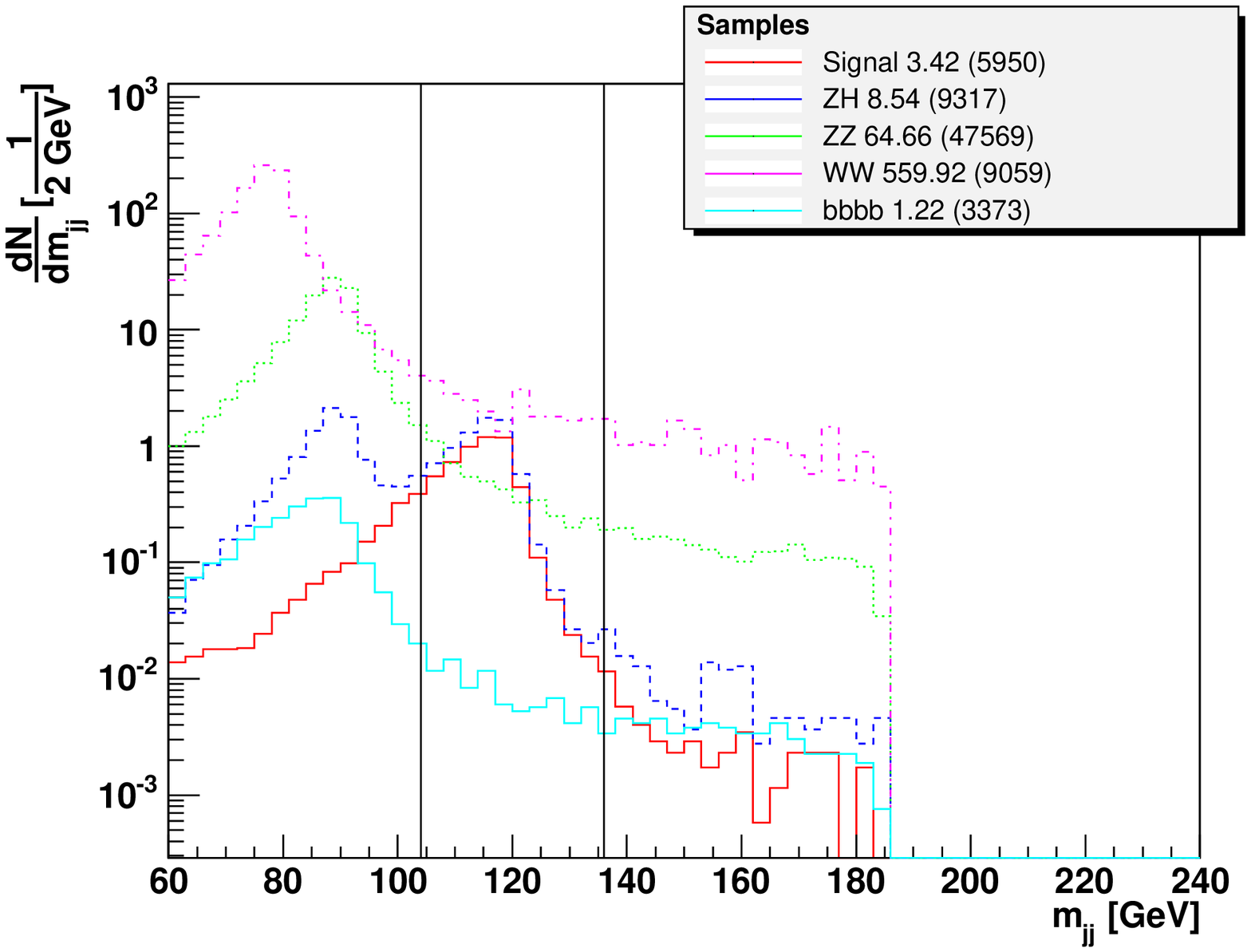, width=8.5cm}
\epsfig{file=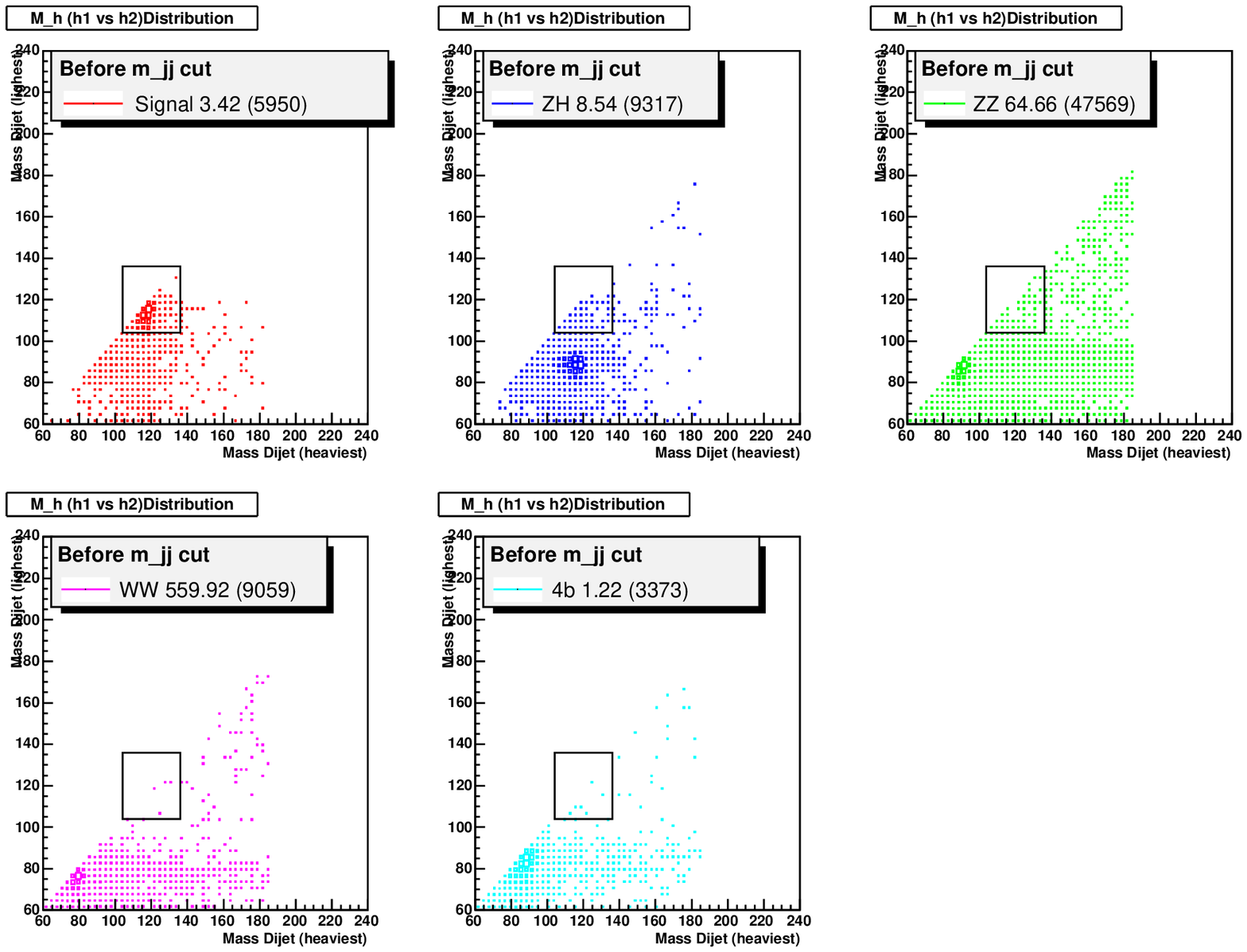, width=8.cm}
\caption{
Distribution of the invariant masses of the dijets used as higgs
candidates in 1 dimension (left) and in 2 dimensions (right). 
Only those which are less than 16~GeV away from the Higgs mass will be 
kept as Higgs candidates (the range kept is within the two vertical
lines on the left plot and inside the boxes on the right plot). The
numbers in this figure correspond to an integrated luminosity of
1~fb$^{-1}$ and the numbers in parentheses correspond to the number of
Monte Carlo events.
}
\label{fig:massh}
\end{center}
\end{figure}

Candidate events for the $e^+e^- \to HH$  process were selected 
among the signal and the background samples by applying the following set of selection criteria. 
The candidates had to have at least 25 tracks with an energy of at least 100 MeV each,
a visible energy of at least 600 GeV, and a transverse momentum of less than 50 GeV. 
A jet finding algorithm (JADE type) was used to count the number of jets in the events 
and those with less than 4 jets at $y_{\mbox{cut} }= 0.004$ were rejected. 
When there were more than 4 jets in an event, they were rearranged to form only 4 jets. 
%Events where % this was not possible, or  
%either jet 
% has an energy lower 5 GeV, or 
%is very forward ($\cos\theta_{\mbox{jet}} > 0.9 $) 
%were rejected. 
Jets were then pair-combined to form all the possible Higgs candidates. 
Only events for which it was possible to form simultaneously 2 pairs 
whose invariant mass was less than 16~GeV away from the Higgs mass were kept. 
Figure~\ref{fig:massh} shows the distribution of the dijets invariant mass before the rejection of the candidates 
more than 16 GeV away from the higgs mass. 
%%%%%%%%%%%%%%%%%%%%%%%%%%%
%  Fig 2
%%%%%%%%%%%%%%%%%%%%%%%%%%
%%%%%%%%%%%%%%%%%%%%%%%%%%
To further improve the purity of the selection a $b$-tagging algorithm\footnote{
When a jet contained three or more tracks with an impact parameter more than
3-$\sigma$ away from the primary vertex, the jet was tagged as a $b$-jet.
}
was used to require at least one of the jets of each Higgs candidate to be $b$-tagged.
%%%%%%%%%%%%%%%%%%%%%%%%%%%
%  Table 1
%%%%%%%%%%%%%%%%%%%%%%%%%%
\begin{table}
\hspace{-1.cm}
\small
\begin{tabular}{| l || l | l | l  |  l |  l |}
\hline
Selection criteria  &          Signal   &         ZZ
&       ZH   &              WW             & bbbb \\ \hline \hline
No cut &  5.772 (10000)    &  206.666 (150000)    &  18.395 (20000)
&  3833.3 (60000)     &  3.779 (10000)  \\ 
$N_{\mbox{tracks}} > 25$    &  5.674 (9831)      &  164.330 (119272)
&  18.202 (19790)    &  2427.1 (37990)    &  3.702 (9796)  \\ 
 $E_{\mbox{vis}} > 600$ GeV  &  5.471 (9479)      &  90.856 (65944)
&  11.287 (12272)    &  1203.8 (18842)    &  2.310 (6111) \\ 
$P_t \le 50$ GeV    &  3.662 (6345)      &  79.912 (58001)      &
8.9160 (9694)      &  939.61 (14707)    &  1.946 (5149)  \\ 
%  3.662 (6345)      &  79.9122 (58001)      &  8.9160 (9694)      &  939.61 (14707)     & $ |P_l| \le 9999$ GeV      \\ 
 $N_{\mbox{jets}} \ge 4$ at $y_{\mbox{cut}} = 0.004$   &  3.481 (6031)
 &  69.682 (50576)      &  8.6308 (9384)      &  644.89 (10094)   &  1.586 (4197) \\ 
%  3.481 (6031)      &  69.6823 (50576)      &  8.6308 (9384)      &  644.89 (10094)     &  $N_{\mbox{jets}} = 4$      \\ 
%  3.481 (6031)      &  69.6823 (50576)      &  8.6308 (9384)      &  644.89 (10094)     &  $E_{\mbox{jet}} > 5$ GeV      \\ 
%$|\cos \theta_j| \le 0.9$     &  2.596 (4498)      &  36.0205 (26144)      &  7.6716 (8341)      &  238.2 (3728)       \\ 
%  2.532 (4386)      &  31.7563 (23049)      &  7.2761 (7911)      &  157.7 (2469)       & 2 jets pairs      \\ 
$|m_{jj} - m_H| \le 16$ GeV    &   2.234 (3870)      &  0.136 (99)
 &  0.174 (190)      &  0.319 (5)   &  2.268 $\times  10^{-3}$  (6)   \\ 
 $b$-tagging     &   1.313 (2275)      &  0.006 (4)      &  0.038 (41)
 &  0.0 (0)    &   3.78 $\times  10^{-4}$ (1)  \\ 
%  1.382 (2394)      &  0.012 (9)               &  0.087 (94)           &  0. (0)      & $|\cos\theta_h| \le 0.99$    \\ 
%  1.382 (2394)      &  0.012 (9)               &  0.087 (94)           &  0. (0)      & $\theta_{\mbox{acop}} \le 180^\circ$      \\ 
\hline
\end{tabular}
\label{Table:cuts}
\caption[]
{
Cut statistics for the signal ($M_S = 2$ TeV) and background event samples with
$M_S=2$~TeV and $M_h = 120$ GeV. The floating point numbers correspond to
the numbers of remaining events per 1 fb$^{-1}$, while the numbers in parentheses represent the
actual numbers of Monte Carlo events surviving the cuts. 
The cross sections include the effects of both 
initial state radiation and beamstrahlung. 
}
\end{table}
%%%%%%%%%%%%%%%%%%%%%%%%%%

%%%%%%%%%%%%%%%%%%%%%%%%%%%
%  Table Mh=160
%%%%%%%%%%%%%%%%%%%%%%%%%%
\begin{table}
\hspace{-.4cm}
\begin{tabular}{| l || l | l | l  |  l |}
\hline
Selection criteria  &          Signal    &         ZZ          &       ZH   &              WW              \\ \hline \hline
No cut & 5.055 (10000)      &  206.666 (150000)      &  16.270 (20000)      &  3833.330 (60000)      \\
$N_{\mbox{tracks}} > 25$    &   4.975 (9841)      &  164.330 (119272)      &  16.185 (19895)      &  2427.137 (37990)      \\
 $E_{\mbox{vis}} > 600$ GeV  & 4.841 (9577)      &  90.856 (65944)      &  10.798 (13273)      &  1203.800 (18842)      \\
$P_t \le 50$ GeV    &  3.235 (6400)      &  79.912 (58001)      &  8.555 (10516)      &  939.613 (14707)      \\
 $N_{\mbox{jets}} \ge 4$ at $y_{\mbox{cut}} = 0.004$   & 3.123 (6177)      &  69.682 (50576)      &  8.308 (10212)      &  644.894 (10094)      \\
$|m_{jj} - m_H| \le 32$ GeV    & 2.506 (4958)      &  0.343 (249)      &  0.106 (130)      &  2.939 (46)      \\
 $b$-tagging     &  1.276 (2524)      &  0.063 (46)      &  0.030 (37)      &  0.000 (0)      \\
\hline
\end{tabular}
\label{Table:cutsMh160}
\caption[]
{
Cut statistics for the signal ($M_S = 2$ TeV) and background event samples with $M_S=2$~TeV and $M_h = 160$ GeV. The floating point numbers correspond to
the numbers of remaining events per 1 fb$^{-1}$, while the numbers in parentheses represent the
actual numbers of Monte Carlo events surviving the cuts. 
The cross sections include the effects of both 
initial state radiation and beamstrahlung. 
}
\end{table}

%%%%%%%%%%%%%%%%%%%%%%%%%%%
%  Table Mh=200
%%%%%%%%%%%%%%%%%%%%%%%%%%
\begin{table}
\hspace{-.4cm}
\begin{tabular}{| l || l | l | l  |  l |}
\hline
Selection criteria  &          Signal    &         ZZ          &       ZH   &              WW              \\ \hline \hline
No cut & 4.219 (10000)      &  206.666 (150000)      &  0.101 (20000)      &  3833.330 (60000)      \\
$N_{\mbox{tracks}} > 25$    &  4.151 (9838)      &  164.330 (119272)      &  0.100 (19758)      &  2427.137 (37990)      \\
 $E_{\mbox{vis}} > 600$ GeV  & 4.052 (9603)      &  90.856 (65944)      &  0.061 (12141)      &  1203.793 (18842)      \\
$P_t \le 50$ GeV    & 2.745 (6507)      &  79.912 (58001)      &  0.048 (9613)      &  939.613 (14707)      \\
 $N_{\mbox{jets}} \ge 4$ at $y_{\mbox{cut}} = 0.004$   & 2.678 (6346)      &  69.682 (50576)      &  0.047 (9325)      &  644.894 (10094)      \\
$|m_{jj} - m_H| \le 32$ GeV    & 1.909 (4524)      &  0.700 (508)      & 4.29$\times 10^{-4}$ (85)      &  8.242 (129)      \\
 $b$-tagging     &  0.842 (1995)      &  0.088 (64)      &
2.17$\times 10^{-4}$ (43)      &  0.127778 (2)      \\

\hline
\end{tabular}
\label{Table:cutsMh200}
\caption[]
{
Cut statistics for the signal ($M_S = 2$ TeV) and background event samples with $M_S = 2$~TeV and $M_h = 200$ GeV. The floating point numbers correspond to
the numbers of remaining events per 1 fb$^{-1}$, while the numbers in parentheses represent the
actual numbers of Monte Carlo events surviving the cuts. 
The cross sections include the effects of both 
initial state radiation and beamstrahlung. 
}
\end{table}

%%%%%%%%%%%%%%%%%%%%%%%%%%%
%  Table M_S
%%%%%%%%%%%%%%%%%%%%%%%%%%
\begin{table}
\hspace{-.4cm}
\begin{tabular}{| l || l | l | l  |  l |}
\hline
Selection criteria  &          Signal ($M_S = 2$ TeV)    &          Signal ($M_S = 2.5$ TeV)    &          Signal ($M_S = 3$ TeV)   \\
\hline \hline
No cut &  5.772 (10000)    & 0.968 (10000) & 0.045 (2012) \\
$N_{\mbox{tracks}} > 25$    &  5.674 (9831)      & 0.952 (9829) &0.221 (9829) \\  
 $E_{\mbox{vis}} > 600$ GeV  &  5.471 (9479)      &  0.918 (9480)&0.213 (9480) \\
$P_t \le 50$ GeV    &  3.662 (6345)      & 0.615 (6351) &  0.143
(6351) \\
 $N_{\mbox{jets}} \ge 4$ at $y_{\mbox{cut}} = 0.004$   &  3.481 (6031)
&  0.584 (6028) & 0.136 (6028)\\
$|m_{jj} - m_H| \le 16$ GeV    &   2.234 (3870)      &  0.374
(3861) & 0.087 (3861)      \\
 $b$-tagging     &   1.313 (2275)      &  0.219 (2264) & 0.051 (2264) \\

\hline
\end{tabular}
\label{Table:cutsMs}
\caption[]
{
Cut statistics for the signal for three differnt values of  $M_S$: 
  $M_S = 2$~TeV , $M_S = 2.5$~TeV  and $M_S = 3$~TeV with $M_h = 120$ GeV. The floating point numbers correspond to
the numbers of remaining events per 1 fb$^{-1}$, while the numbers in parentheses represent the
actual numbers of Monte Carlo events surviving the cuts. 
The cross sections include the effects of both 
initial state radiation and beamstrahlung. These
number of event scale like $\frac{1}{M_S^8}$, as expected from Eq.(\ref{eq:diffcrosssection}).
}
\end{table}

Tables~\ref{Table:cuts}, \ref{Table:cutsMh160} and~\ref{Table:cutsMh200} summarizes the number of surviving events 
per 1 fb$^{-1}$ at each step of the selection cuts
for both  the signal and background event samples for different values
of the Higgs mass ($M_h = 120$ GeV, $M_h = 160$ GeV and $M_h = 200$
GeV).
The table~\ref{Table:cutsMs} confirms that the signal's differential cross section
scales like $\frac{1}{M_S^8}$, as calculated in Eq.(\ref{eq:diffcrosssection}).
These tables show that the above selection criteria reject 
almost all the backgrounds,
while keeping a reasonable detection efficiency around 20\% 
for the signal process. 
For the sample set of parameters, $M_S=2~\mbox{TeV}$ and $m_H=120~\mbox{GeV}$,
this corresponds to about 700 events at $\sqrt{s} = 1$ TeV for an integrated luminosity of
500 fb$^{-1}$. 
For the processes 
$e^+ e^- \rightarrow ZZ $  and $e^+ e^- \rightarrow W^+W^-$  only the Standard Model 
contribution has been considered. 
Since these processes could also be mediated by virtual KK gravitons and hence be enhanced, 
the $e^+e^- \to ZZ$ process would have become the dominant background.
Even so, the number of events surviving the selection criteria would remain
at least one order of magnitude lower than our signal.
On the other hand, the $e^+e^- \to ZH$ background would stay the same, 
since there would be no virtual KK graviton contribution.
\label{sec:zzkk}

\subsection{Reconstruction of Angular Distribution}

\begin{figure}
\begin{center}
\epsfig{file=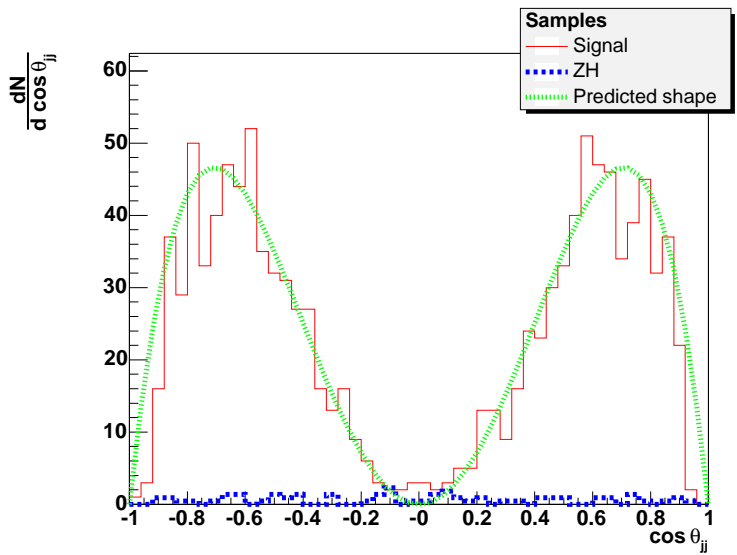, width=10cm}
\caption{
Distribution of the $\cos\theta_{\mbox{jet}}$ of the signal events (in red) and the ZH background (in blue) after application of the selection criteria. The numbers in this figure correspond to an integrated luminosity of 500~fb$^{-1}$.
}
\label{fig:costheta}
\end{center}
\end{figure}

Using the selected Higgs pair candidates, we can now look at the angular distribution of the
reconstructed Higgs pairs.
Figure~\ref{fig:costheta} shows a typical distribution of the reconstructed $\cos\theta$ 
of the selected events compared to the shape expected from Eq.~\ref{eqn:dsigma} (green dotted curve). The discrepancy near $\cos\theta_{\mbox{jj}} = \pm 1$ comes from the limited acceptance of the detector near the beam pipe. 
%%%%%%%%%%%%%%%%%%%%%%%%%%%
%  Fig 3
%%%%%%%%%%%%%%%%%%%%%%%%%%
%%%%%%%%%%%%%%%%%%%%%%%%%%
The reconstructed $\cos\theta$ distribution clearly exhibits the characteristic spin-2 behavior,
demonstrating the possibility of the test of KK graviton spin exchanged in the $s$-channel.
%\footnote{
%The distribution does not go down to zero at $\cos\theta = 0, \pm 1$. This is mostly due to the
%finite angular resolution for jet reconstruction.
%}

\section{Summary and Conclusions}
We have derived the cross section formula for the process $e^+e^- \to HH$ via
$s$-channel KK graviton exchange.
We have found that the cross section at $\sqrt{s} = 1$ TeV is of the same order of 
magnitude as that of the associated higgs production cross section, 
$e^+ e^- \rightarrow Z H$ in the standard model,
for a sample parameter set, $M_S=2~\mbox{TeV}$ and
$m_H=120~\mbox{GeV}$.

It should be stressed that the differential cross section gives a characteristic behavior reflecting the
spin 2 nature of the intermediate KK gravitons.
The cross section formula was then used to carry out Monte Carlo simulations to test
the feasibility of measuring this angular distribution.
We have shown that, for our sample set of parameters,
we can obtain about 700 Higgs pair events at $\sqrt{s} = 1$ TeV,
given an integrated luminosity of 500 fb$^{-1}$, with essentially no
Standard Model backgrounds. 
Even if $M_S$ is higher the $2$ TeV but remains below 2500 GeV, we will be able to accumulate
more than 100 events at $\sqrt{s} = 1$TeV and observe the characteristic
angular distribution. Conversely, if $M_S=4$TeV, we would need
to accumulate 500 $fb^{-1}$ at $\sqrt{s} = 3.4$~TeV or higher to be
able to accumulate 100~Higgs pair events.

Using this very clean sample, we have then demonstrated that we can indeed measure
the characteristic angular distribution of the Higgs pair production
via the KK graviton exchange.

Finally, as mentionned in section~\ref{sec:zzkk}, 
the processes $e^+ e^- \rightarrow ZZ$  and $e^+ e^- \rightarrow W^+W^-$  
can also be mediated by virtual KK gravitons. 
This has not been taken into account in our simulations but our
estimations shows that in the worse case the $ZZ$ cross section would
be enhanced 4 times and the $WW$ cross section 40\% higher, thus the
background coming from these processes would remain very small.

The  processes $e^+ e^- \rightarrow ZZ$  and $e^+ e^- \rightarrow W^+W^-$  
mediated by virtual KK gravitons are
 certainly 
worth investigating for the linear collider studies of extra dimension scenarios~\cite{DFO}.
%and will be done soon. 

\vspace{1cm}
%%%%%%%%%%%%%%%%%%%%%%%%%%%%%%%%%%%%%%%%%%%%%%%%%%%%%%%%%%%%%%%%%%%%%%%%%%%%%%%
%                             Acknowledgments                                 %
%%%%%%%%%%%%%%%%%%%%%%%%%%%%%%%%%%%%%%%%%%%%%%%%%%%%%%%%%%%%%%%%%%%%%%%%%%%%%%%

\begin{center}
{\bf Acknowledgments}
\end{center}
The authors would like to thank all the members of the new physics
subgroup of the LC physics study group in Japan.
Among them Shinji Komine deserves special mention for his
contribution in the early stage of this work.
One of the authors (ND) would also like to thank JSPS 
for funding his stay in Japan under contract P02794.
This work is partially supported by JSPS-CAS Scientific Cooperation 
Program under the Core University System and by Japan-Europe (UK) 
Research Cooperative Program of JSPS.

\vspace{1cm}
%%%%%%%%%%%%%%%%%%%%%%%%%%%%%%%%%%%%%%%%%%%%%%%%%%%%%%%%%%%%%%%%%%%%%%%%%%%%%%%
%                                  References                                 %
%%%%%%%%%%%%%%%%%%%%%%%%%%%%%%%%%%%%%%%%%%%%%%%%%%%%%%%%%%%%%%%%%%%%%%%%%%%%%%%

\end{document}